\documentclass[10pt]{article}
\usepackage[cp866]{inputenc}
\begin{document}
\bigskip

\centerline {{\Large\bf Two types of conservation laws. Connection }}
\centerline {{\Large\bf of physical fields with material systems.}}
\centerline {{\Large\bf Peculiarities of field theories}}
\centerline {\it L.~I. Petrova}
 
\renewcommand{\abstractname}{Abstract}
\begin{abstract}

Historically it happen so that in branches of physics connected with
field theory and of physics of material systems (continuous media)
the concept of "conservation laws" has a different meaning.
In field theory "conservation laws" are those that claim the
existence of conservative physical quantities or objects. These are
conservation laws for physical fields. In contrast to that in physics
(and mechanics) of material systems the concept of ``conservation laws"
relates to conservation laws for energy, linear momentum, angular 
momentum, and mass that establish the balance between the change 
of physical quantities and external action.

In the paper presented it is proved that there exist a connection between 
of conservation laws for physical fields and those for material systems. 
This points to the fact that physical fields are connected with material 
systems.

Such results has an unique significance for field theories. This
enables one to substantiate many basic principles of field
theories, such as, for example, the unity of existing field
theories and the causality. The specific feature of field theory
equations, namely, their connection to the equations for material
systems, is elicited.

Such results have been obtained by using skew-symmetric differential 
forms, which reflect the properties of conservation laws.

(It appears to be possible to obtain radically new results due to using
skew-symmetric differential form, which, in contrast to exterior forms,
are defined on nonintegrable manifolds and possesses evolutionary 
properties. The mathematical apparatus of evolutionary forms includes 
nontraditional elements, that are, nonidentical relations and degenerate 
transformations. This enables one to describe the mechanism of arising 
and generating the structures that is
impossible to perform within the framework of any existing mathematical 
formalisms. The existence of such skew-symmetric differential forms has 
been established by the author while studying the stability problems.)

\end{abstract}

\section{Meaning of the concept of ``conservation laws"}
Due to the development of science the concept of ``conservation laws"
in thermodynamics, physics and mechanics contains different meanings.

In branches of physics related to field theory and in theoretical
mechanics ``the conservation laws" are those according to which there
exist conservative physical quantities or objects. These are the
conservation laws that were called above as ``exact" ones.

In mechanics and physics of continuous media the concept of
``conservation laws" relates to conservation laws for energy,
linear momentum, angular momentum, and mass, which establish the balance
between the change of physical quantities and external action. These are
balance conservation laws.

In thermodynamics conservation laws are associated with
the principles of thermodynamics.

Thus, the concept of ``conservation laws" is connected with exact
conservation laws, i.e. balance conservation laws and some regularities
expressed using the principles of thermodynamics.

Below it will be shown that the balance conservation laws are those for
continuous media (material systems). The exact conservation laws are
conservation laws for physical fields.

In addition it will be shown that the balance and exact conservation
laws are related to each other. (The principles of thermodynamics
integrate two balance conservation laws, namely, the balance
conservation law for energy and that for linear momentum.)

The connection between conservation laws for physical fields and
those for material systems discloses the peculiarity of physical
fields, namely, their connection to material systems.

This has an unique importance for field theory as well. It will be shown
that the equations of field theory, which are based on the conservation 
laws for physical fields, are connected with the equations that 
describe conservation laws for material systems.                 
This provides answers to many problems of existing field theories.

The mathematical apparatus of skew-symmetric differential forms allows 
to describe conservations laws and their peculiarities.

\section{Closed exterior skew-symmetric differential forms:
Conservation laws for physical fields. Physical structures}

\subsection{Conservation laws for physical fields. Physical structures}

As it was already pointed out, the conservation laws for physical fields
are those that state the existence of conservative physical quantities 
or objects.
{\footnotesize [The physical fields [1] are
a special form of substance, they are carriers of various
interactions such as electromagnetic, gravitational, wave, nuclear and
other types of interactions.]}

The conservation laws for physical fields are described by closed 
exterior differential forms [2,3].

From the closure condition of closed exterior differential forms 
$d\theta^p$ 
$$
d\theta^p=0\eqno(1)
$$
one can see that the closed form is a conservative
quantity. This means that this form can correspond to conservation law,
namely, to some conservative physical quantity.

If the form is closed on a pseudostructure only (closed inexact form),
the closure conditions is written as
$$d_\pi\theta^p=0\eqno(2)$$
$$d_\pi{}^*\theta^p=0\eqno(3)$$
where ${}^*\theta^p$ is the dual form.

From conditions (2) and (3) one can see that the form
closed on pseudostructure is a conservative object, namely, this
quantity conserves on pseudostructure. This can also correspond to
some conservation law, i.e. to conservative object.

The closed (inexact) exterior form and dual form describe the
differential-geometrical structure. It is evident that such a structure
(a conservative object) must correspond to exact conservation law.
Such structures (pseudostructures with a conservative physical
quantity), which correspond to exact conservation law, describe
the physical structures that made up physical fields.

The problem of how physical structures arise and how physical fields
are formatted will be discussed below.

The equations for physical structures ($d_{\pi }\,\theta ^p\,=\,0$,
$d_{\pi }\,^*\theta ^p\,=\,0$) turn out to coincide with the mathematical
expression for exact conservation law.

The mathematical expression for exact conservation law and
its relation to physical fields can be schematically written
in the following manner
$$
\def\\{\vphantom{d_\pi}}
\cases{d_\pi \theta^p=0\cr d_\pi {}^{*\mskip-2mu}\theta^p=0\cr}\quad
\mapsto\quad
\cases{\\\theta^p\cr \\{}^{*\mskip-2mu}\theta^p\cr}\quad\hbox{---}\quad
\hbox{physical structures}\quad\mapsto\quad\hbox{physical fields}
$$

It is seen that the exact conservation law is that for physical fields.

(It should be emphasized that the closed {\it inexact } forms correspond
to the physical structures that made up physical fields. The {\it exact}
forms correspond to the material system {\it elements}.)

\subsection{Closed inexact exterior forms as the basis of field theories}

The field theories, i.e. the theories that describe physical fields,
are based on the properties of closed inexact exterior differential and 
dual forms that correspond to exact conservation laws.

The invariant properties of closed exterior differential forms
reveal themselves explicitly or implicitly in practically all formalisms  
of field theories such as the Hamilton formalism, tensor approaches, 
group methods, quantum mechanic equations, the Yang-Mills theory and so 
on.

The nondegenerate transformations of field theories are those of closed 
exterior forms.

Since the closed form is a differential (a total one
if the form is exact, or an interior one on the pseudostructure if the
form is inexact), it is obvious that the closed form turns out to be
invariant under all transformations that conserve the differential.
The unitary, tangent, canonical and gradient transformations
and so on are examples of such transformations of closed exterior forms.
{\it These are gauge transformations of field theories}.
The gauge transformations for spinor, scalar, vector and tensor fields 
are respectively transformations of closed ($0$-form), ($1$-form), 
($2$-form) and ($3$-form).

The gauge, i.e. interior, symmetries of field theories (which correspond
to gauge transformations) are symmetries of closed exterior forms. 
Exterior symmetries of the field theory equations are those of closed 
dual forms.

Operators of the field theory are connected to nondegenerate 
transformations of closed exterior forms.
In terms of operators $d$ (exterior differential), $\delta$ 
(the operator of transformation that converts the form of degree
$p+1$ into the form of degree $p$), $\delta '$ (for cotangent 
transformations), $\Delta $ (for the transformation $d\delta-\delta d$), 
$\delta$ (for transformation $d\delta'-\delta' d$) one can write 
the operators of field theory equations. In terms of these operators, 
which act to exterior forms, it is possible to write the Green, 
d'Alambert, Laplace operators, as well as the operator
of canonical transformation [4].

One can make sure that all existing field theories were built on the 
postulates of invariance and covariance, which are closure conditions 
for exterior and dual forms.

And there is the following correspondence.

-Closed exterior forms of zero degree correspond to quantum mechanics.

-The Hamilton formalism bases on the properties of closed exterior and
dual forms of first degree.

-The properties of closed exterior and dual forms of second degree are
at the basis of the equations of electromagnetic field.

-The closure conditions of exterior and dual forms of third degree form
the basis of equations for gravitational field.

Closed inexact exterior or dual forms are solutions of the
field theory equations.

One can see that field theory equations as well as nondegenerate 
transformations of field theories are connected with closed
exterior forms of a certain degree.
This enables one to introduce the classification of physical fields and
interactions in degrees of closed exterior forms. (If to denote the
degree of closed exterior forms by $k$, then $k=0$ corresponds to
strong interaction, $k=1$ does to weak one, $k=2$ does to
electromagnetic one, and $k=3$ corresponds to gravitational interaction.) 

Such a classification shows that there exists an internal connection
between field theories, which describe physical fields of various
types. It is evident that the degree of closed exterior forms is a
parameter that integrates field theories into unified field theory.

A significance of exterior differential forms for field theories
consists in the fact that they disclose the properties that are common
for all field theories and physical fields irrespective of their
specific type. This is a step to building a unified field theory.

However, the theory of exterior forms cannot solve the
problem of justifying the principles that are at the basis of field
theories. To solve these problems, one has to answer the following
questions:

(a) how are obtained the closed and dual forms that made up physical 
structures and on the properties of which the field theories are based;

(b) what is a physical meaning of such a parameter like a degree of 
closed exterior forms, which unite field theories;

(c) by what are conditioned the symmetries of closed exterior and dual 
forms that are assigned to internal and external symmetries of field 
theories;

(d) with what are connected the transformations of closed exterior forms, 
that is, with what are connected the gauge transformations in field 
theories;

(e) how to explain a discrete realization of closed (inexact) exterior
forms that could disclose the quantum nature of field theories.

Within only framework of exterior differential forms it is
impossible to answer these questions. Below it will be shown that
closed exterior forms, whose properties lie at the basis of field
theories, are realized from evolutionary forms obtained from the
equations that describe conservation laws for material systems.
This enables one to understand the basic principles of field
theories, namely, their connections with the equations for
material systems, and answer the questions posed up.

\section{Evolutionary differential forms: Conservation laws for material 
systems. Noncommutativity of conservation laws for material systems.}

The properties of balance conservation laws, as well as the
properties of exact conservation laws, are described by
skew-symmetric differential forms. But these skew-symmetric forms,
as contrasted to exterior forms, are defined on nonintegrable
manifolds and possess the evolutionary properties [5]. {\footnotesize
[As examples of manifolds, with which the evolutionary forms are
connected, are the tangent manifolds of differential equations
that describe any processes, the Lagrangian manifolds, the
manifolds made up by trajectories of material system elements
(particles), which are obtained while describing the evolutionary
processes in material media, and others.]} 

(The evolutionary forms have an unique peculiarity. From 
the evolutionry form obtained from the equations describing the balance 
conservation laws for material systems (continuous media) the closed 
exterior forms, which describe the conservation laws for physical 
fields, are realized. This means that physical fields are connected 
with material systems.)

\subsection{Conservation laws for material systems}

The conservation laws for material systems (material media) are
conservation laws of energy, linear momentum, angular momentum, and 
mass. In contrast to conservation laws for physical fields these 
conservation laws are balance ones.

{\footnotesize [The material system is a variety of elements which have
internal structure, move and interact to one another. Examples of 
elements that constitute material system are electrons, protons, 
neutrons, atoms, fluid particles, cosmic objects and others. 
As examples of material systems it may be thermodynamic, gas dynamical, 
cosmic systems, systems of elementary particles (pointed above) and 
others. The physical vacuum in its properties may be regarded as 
an analogue of material system that generates some physical fields. 
Any material media are such material systems]}

In mechanics and physics of material systems (of continuous media)
the equations of balance conservation laws are used for
description of physical quantities, which specify the behavior of
material systems. However, it turns out that the role of these
equations is much wider. They describe evolutionary processes in
material systems that are accompanied by an origin of physical
structures, from which physical fields are formatted.

The equations of balance conservation laws are differential
or integral equations [6-8]. {\footnotesize (The equations of mechanics
and physics of continuous media such as the Euler and Navier-Stokes
equations [7] are examples.)}

The functions sought being solutions to equations of material
media are usually functions which relate to such physical
quantities like a particle velocity (of elements), temperature or
energy, pressure and density. Since these functions relate to one
material system, it has to exist a connection between them. This
connection is described by some state functional (from which the
state function is obtained). From the equations of balance
conservation laws one gets the relation for state functional that,
as it will be shown below, is of great significance both in
mathematical physics, which describes material systems, and in
field theories, which describe physical fields.

\subsection{Analysis of the equations of balance conservation laws. 
Evolutionary relation.}

It appears that, even without a knowledge of concrete form
of the balance conservation law equations, with the help of
skew-symmetric differential forms one can see specific features of
these equations that elucidate the properties of balance conservation
laws and their role in evolutionary processes.

The functional properties of equations or sets of equations depend on
whether or not the derivatives of differential equations or of the
equation in the set of differential equations are conjugated.
{\footnotesize [The necessity of studying the conjugacy of equations
describing any process has a physical meaning. If these equations
(or derivatives with respect to different variables) be not conjugated,
the solutions to corresponding equations prove to be noninvariant, that 
is, they are functionals rather then functions. The realization of
the conditions (while varying variables), under which the equations
become conjugated ones, leads to that the relevant solution becomes
invariant. It will be shown below that the transition to invariant
solution describes the mechanism of evolutionary transition from one
quality to another, which leads to emergence of differential-geometrical
structures]}.

Equations are conjugate if they can be contracted into identical
relations for differential, i.e. for a closed form.

Let us analyze the equations that describe the balance conservation
laws for energy and linear momentum.

We introduce two frames of reference: the first is an inertial one
(this frame of reference is not connected with material system), and
the second is an accompanying one (this system is connected with the 
manifold built by trajectories of material system elements). The energy 
equation in the inertial frame of reference can be reduced to the form:
$$
\frac{D\psi}{Dt}=A \eqno(4)
$$
where $D/Dt$ is the total derivative with respect to time, $\psi $ is 
the functional of the state that specifies material system, $A$ is the 
quantity that depends on specific features of the material system and 
on external energy actions onto the system. \{The action functional, 
entropy, wave 
function can be regarded as examples of the functional $\psi $. Thus, 
the equation for energy expressed in terms of the action functional 
$S$ has a similar form: $DS/Dt\,=\,L$, where $\psi \,=\,S$, $A\,=\,L$ 
is the Lagrange function. In mechanics of continuous media the equation 
for energy of ideal gas can be expressed in the form [6]: $Ds/Dt\,=\,0$, 
where $s$ is entropy. In this case $\psi \,=\,s$, $A\,=\,0$. It is worth 
noting that the examples presented demonstrate that the action 
functional and entropy play the same role.\}

In accompanying frame of reference the total derivative with respect to
time is converted into a derivative along trajectory.
Equation (4) is now written in the form
$$
{{\partial \psi }\over {\partial \xi ^1}}\,=\,A_1 \eqno(5)
$$
here $\xi^1$ is the coordinate along trajectory ($A_1=A$).

In a similar manner, in accompanying frame of reference the equation 
for linear momentum appears to be reduced to the equation of the form
$$
{{\partial \psi}\over {\partial \xi^{\nu }}}\,=\,A_{\nu },\quad \nu \,=\,2,\,...\eqno(6)
$$
where $\xi ^{\nu }$ are the coordinates in the direction normal to
trajectory, $A_{\nu }$ are the quantities that depend on specific
features of the system and external force actions.

Eqs. (5), (6) can be convoluted into the relation
$$
d\psi\,=\,A_{\mu }\,d\xi ^{\mu },\quad (\mu\,=\,1,\,\nu )\eqno(7)
$$
where $d\psi $ is the differential
expression $d\psi\,=\,(\partial \psi /\partial \xi ^{\mu })d\xi ^{\mu }$. 

Relation (7) can be written as
$$
d\psi \,=\,\omega \eqno(8)
$$
here $\omega \,=\,A_{\mu }\,d\xi ^{\mu }$ is the differential form of
first degree.

Since the balance conservation laws are evolutionary ones, the relation
obtained is also an evolutionary relation.

Relation (8) was obtained from the equations of balance conservation
laws for energy and linear momentum. In this relation the form $\omega $
is that of first degree. If the equations of balance conservation
laws for angular momentum be added to the equations for energy and
linear momentum, this form in evolutionary relation will be a form
of second degree. And in combination with the equation of balance
conservation law of mass this form will be a form of degree 3.

Thus, in general case the evolutionary relation can be written as
$$
d\psi \,=\,\omega^p \eqno(9)
$$
where the form degree  $p$ takes the values $p\,=\,0,1,2,3$.
(The evolutionary
relation for $p\,=\,0$ is similar to that in differential forms, and it 
was obtained from interaction of energy and time.)

In relation (8) the form $\psi$ is a form of zero degree. And in
relation (9) the form $\psi$ is a form of $(p-1)$ degree.

\subsection{Nonidentity of evolutionary relation: Noncommutativity of 
balance conservation laws.}

Let us show that {\it the evolutionary relation  obtained from the
equation of balance conservation laws proves to be nonidentical one}.

To do so we shall analyze relation (8).

The relation may be identical one if this is the relation between 
measurable (invariant) quantities or between observable (metric) objects, 
in other words, between quantities or objects that are comparable.

In the left-hand side of evolutionary relation (8) there is a
differential that is a closed form. This form is an invariant object.
The right-hand side of relation (8) involves the differential form
$\omega$, that is not an invariant object because in real processes,
as it is shown below, this form proves to be unclosed.

For the form be closed the differential of the form or its commutator
must be equal to zero (the elements of the form differential are equal
to components of its commutator).

Let us consider the commutator of the
form $\omega \,=\,A_{\mu }d\xi ^{\mu }$. The components
of commutator of such a form can be written as follows:
$$
K_{\alpha \beta }\,=\,\left ({{\partial A_{\beta }}\over {\partial \xi ^{\alpha }}}\,-\,
{{\partial A_{\alpha }}\over {\partial \xi ^{\beta }}}\right )
$$
(here the term  connected with a nonintegrability of the manifold
has not yet been taken into account).

The coefficients $A_{\mu }$ of the form $\omega $ have been obtained 
either from the equation of balance conservation law for energy or from 
that for linear momentum. This means that in the first case the 
coefficients depend on energetic action and in the second case they 
depend on force action. In actual processes energetic and force actions 
have different nature and appear to be inconsistent. The commutator of 
the form $\omega $ made up of derivatives of such coefficients is 
nonzero. This means that the differential of the form $\omega $
is nonzero as well. Thus, the form $\omega$ proves to be unclosed and 
is not a measurable quantity.

This means that the evolutionary relation involves an unmeasurable
term. Such a relation cannot be identical one. (In the left-hand side
of this relation it stands a differential, whereas in the right-hand
side it stands an unclosed form that is not a differential.)
[Nonidentical relation was analyzed in paper J.L.Synge "Tensorial
Methods in Dynamics" (1936). And yet it was allowed
a possibility to use the sign of equality in nonidentical relation.]

{\footnotesize [The nonidentity of evolutionary relation does not mean 
that the mathematical description of physical processes is not perfectly
exact. The nonidentity of the relation means that the derivatives,
whose values correspond to real values in physical processes,
cannot be consistent (their mixed derivatives cannot be commutative;
the commutator is nonzero) because they are obtained at the expense
of external action and are unmeasurable quantities.]}

The nonidentity of evolutionary relation means that the equations
of balance conservation laws turn out to be nonconjugated (and
hence, nonintegrable: one  cannot convolute them into an identical
relation and obtain a differential). And this indicates that the
balance conservation laws are noncommutative.

As noted above, each balance conservation law depends on relevant
action (that the material system is subjected to). So, the conservation
law for energy depends on energetic action, the conservation law
for linear momentum depends on force action, and so on. In actual
processes the energetic and the force actions have different nature,
and this is the cause of noncommutativity of balance conservation
laws. The balance conservation laws that depend on actions of different
nature cannot be commutative.

Thus, the nonidentity of evolutionary relation (see, formulas
(8), (9)) means that the balance conservation law equations are
inconsistent. And this indicates that the balance conservation laws are
noncommutative. (If the balance conservation laws be commutative,
the equations would be consistent and the evolutionary relation would
be identical).

{\footnotesize \{The nonidentity of evolutionary relation are
connected with the differential form $\omega^p $ that enters into this
relation. This is a skew-symmetric form with the basis, 
in contrast to the basis of exterior form, that is a deforming 
(nonintegrable) manifold. The peculiarity of skew-symmetric forms
defined on such manifold is the fact that their differential
depends on the basis. The commutator of such form includes the
term that is connected with differentiating the basis. This can
be demonstrated by an example of the first-degree skew-symmetric form.

Let us consider the first-degree form
$\omega=a_\alpha dx^\alpha$. The differential of this form can
be written as $d\omega=K_{\alpha\beta}dx^\alpha dx^\beta$, where
$K_{\alpha\beta}=a_{\beta;\alpha}- a_{\alpha;\beta}$ are
the components of commutator of the form $\omega$, and
$a_{\beta; \alpha}$, $a_{\alpha;\beta}$ are the covariant
derivatives. If we express the covariant derivatives in terms of
connectedness (if it is possible), they can be written
as $a_{\beta;\alpha}=\partial a_\beta/\partial
x^\alpha+\Gamma^\sigma_{\beta\alpha}a_\sigma$, where the first
term results from differentiating the form coefficients, and the
second term results from differentiating the basis. If we substitute
the expressions for covariant derivatives into the formula for
commutator components, we obtain the following expression
for the commutator components of the form $\omega$:
$$
K_{\alpha\beta}=\left(\frac{\partial a_\beta}{\partial
x^\alpha}-\frac{\partial a_\alpha}{\partial
x^\beta}\right)+(\Gamma^\sigma_{\beta\alpha}-
\Gamma^\sigma_{\alpha\beta})a_\sigma
$$
Here the expressions
$(\Gamma^\sigma_{\beta\alpha}-\Gamma^\sigma_{\alpha\beta})$
entered into the second term are just components of commutator of
the first-degree metric form that specifies the manifold
deformation and hence is nonzero. (In commutator of the
exterior form, which is defined on differentiable manifold
the second term is not present.) [It is well-known that
the metric form commutators of the first-, second- and third
degrees specifies, respectively,  torsion, rotation and
curvature.

The skew-symmetric differential forms defined on nonintegrable manifolds 
are evolutionary forms. They are obtained while describing some 
processes. \}}

\bigskip
In the next section by the analysis of nonidentical evolutionary
relation it will be shown that the noncommutativity of balance
conservation laws is a driving force of evolutionary processes that 
proceed in material systems and are accompanied by generation of 
physical structures.

\section {Physical significance of nonidentical evolutionary relation:
Relation of conservation laws for physical fields to conservation
laws for material systems. Relation of physical fields to material
systems}

\subsection{Nonidentity of evolutionary relation: nonequilibrium state
of material system}

The noncommutativity of balance conservation laws is a characteristics
of the state of material system. This is reflected by the evolutionary
relation.

In the left-hand side of evolutionary relation (see, (8, 9)) there
is the functional expression $d\psi$ that determines the state of
material system.

It is evident that if the balance conservation laws be
commutative, the evolutionary relation would be identical and from
that it would be possible to get the differential $d\psi $ and
find the state function $\psi $, this would indicate  that
material system is in equilibrium state. However, as it has been
shown, in real processes the balance conservation laws are
noncommutative. The evolutionary relation is not identical and
from this relation one cannot get the differential $d\psi $. The
absence of differential $d\psi$ points to the fact that the
material system state is nonequilibrium. (As it will be shown
below, the interior (on pseudostructure) differential $d\psi$ can
be realized and this will correspond to locally equilibrium
state.)

The evolutionary relation possesses one more peculiarity,
namely, this relation is a selfvarying relation. (The evolutionary form
entering into this relation is defined on deforming manifold
made up by trajectories of material system elements. This means
that the evolutionary form basis varies. In turn,
this leads to variation of evolutionary form, and the process
of intervariation of evolutionary form and of the basis is
repeated.)

Selfvariation of nonidentical evolutionary relation points to the
fact that the nonequilibrium state of material system turns out
to be selfvarying. State of material system changes but remains
nonequilibrium during this process.

It is evident that selfvariation of nonequilibrium state of material
system proceeds under the action of internal force whose quantity is
described by the commutator of unclosed evolutionary
form $\omega^p $. (If the evolutionary form commutator be zero,
the evolutionary relation would be identical, and this would point to
the equilibrium state, i.e. the absence of internal forces.) Everything
that gives a contribution into the evolutionary form commutator leads
to emergence of internal force.

Is the transition from nonequilibrium state to equilibrium one
possible?

Since the evolutionary form is unclosed, the evolutionary relation
cannot be identical. This means that the nonequilibrium state of 
material system holds.
However, from the evolutionary relation it can be obtained identical
relation on some structure (more exactly, on pseudostructure). And this 
will correspond to transition of material system to locally 
equilibrium state.

To obtain an identical relation from evolutionary nonidentical relation,
it is necessary that a closed exterior differential form should be 
derived from evolutionary differential form that is included into 
nonidentical relation. A transition from evolutionary form (whose 
differential is {\it nonzero}) to  closed exterior forms (whose 
differential is {\it zero}) is possible only as degenerate 
transformation, that is, a transformation that does not conserve 
the differential.

Such a transformation is possible under realization of appropriate 
conditions.

The conditions of degenerate transformation are connected with
symmetries caused by degrees of freedom of material system. The
translational degrees of freedom, internal degrees of freedom of
the system elements, and so on can be examples of such degrees of
freedom. {\footnotesize [To the degenerate transformation it must
correspond a vanishing of some functional expressions, such as
Jacobians, determinants, the Poisson brackets, residues and
others. Vanishing of these functional expressions is a closure
condition for dual form. And it should be emphasize once more that
the degenerate transformation is realized as a transition from
accompanying noninertial frame of reference to locally inertial
system. The evolutionary form and nonidentical evolutionary
relation are defined in noninertial frame of reference (deforming
manifold). But the closed exterior form and the identical relation
are obtained with respect to the locally-inertial frame of
reference (pseudostructure). Mathematically this is described as a
transition from noninertial frame of reference to inertial one.]}

The conditions of degenerate transformation can be realized
under selfvariation of evolutionary relation.

The realization of the conditions of degenerate transformation leads to 
realization of pseudostructure $\pi$ (the closed dual form) and 
formatting the closed inexact form $\omega_\pi$, whose closure 
conditions have the form 
$$d_\pi \omega^p=0,  d_\pi{}^*\omega^p=0 \eqno(10)$$
[Pseudostructures specify the integral surfaces: the characteristics
(the determinant of coefficients at the normal derivatives vanishes),
the singular points (Jacobian is equal to zero), the envelopes of 
characteristics of Eulers' equations and so on.]
{\footnotesize \{\{Cohomology (de Rham cohomology, singular 
cohomology [9]), sections of cotangent bundles, integral and potential 
surfaces and so on may be regarded as examples of pseudostructures.
The eikonal surfaces correspond to pseudostructures\}}

On the pseudostructure $\pi$ from the evolutionary form $\omega^p$ it 
arises (under degenerate transformation) the closed inexact exterior 
form $\omega_\pi^p$ and from evolutionary relation (9) it is obtained 
the relation
$$
d_\pi\psi=\omega_\pi^p\eqno(11)
$$
which proves to be an identical relation  since the closed inexact form
$\omega_\pi^p$ is a differential (interior on pseudostructure).

{\it From identical relation one can obtain the state differential
$d_\pi\psi $ and find the state function, and this points to that
the material system state is a local equilibrium state.}

Thus, from the properties of nonidentical evolutionary relation and 
those of evolutionary form one can see that under realization of the
additional condition (which is a condition of degenerate transformation) 
the transition of material system state from nonequilibrium to locally 
equilibrium state can be realized.

But in this case the total state of material system turns out
to be nonequilibrium because the evolutionary relation itself remains
to be nonidentical one. The equilibrium state is realized only
locally since the state differential is interior one defined
exclusively on pseudostructure.

{\footnotesize [Here one can see the simultaneous presence of
identical and nonidentical relation. This peculiarity discloses
the duality of the quantity $\psi $. From identical relation one
can find the differential $d\psi$ on pseudostructure and after 
integrating obtain the value $\psi $. In this case  $\psi $ 
on pseudostructure becomes the state function.
However, since the evolutionary relation remains to be
nonidentical, this means that the total differential $d\psi$ is not
defined and $\psi $ occurs to be a certain functional. Such a
duality one can trace by the example of entropy. Entropy can be
functional, because it is impossible to find the entropy
differential (total) from nonidentical relation, and be
simultaneously the state function on pseudostructure if the
interior (on pseudostructure) differential of entropy is
realized.]}

Under realization of new additional conditions a new identical relation
can be obtained. As a result, the nonidentical evolutionary relation
can generate identical relations.

\subsection {Realization of physical structures that form physical 
fields. Connection of physical structures with material systems.}

The emergence of the closed (on pseudostructure) inexact exterior form
$\omega_\pi^p$ points to origination of physical structure.
The closure conditions (10) for exterior inexact form correspond to
conservation law, i.e. to existence of conservative on pseudostructure
quantity, and describe a differential-geometrical structure. These
are such structures (pseudostructures with conservative
quantities) that are physical structures, from which physical
fields are formatted.

The transition from nonidentical relation (9) obtained from balance 
conservation laws equations to identical relation (11) means the 
following.
Firstly, an emergence of closed (on pseudostructure) inexact exterior 
form (right-hand side
of relation (11)) points to origination of physical structure.
And, secondly, the existence of state differential (left-hand side
of relation (11) points to  transition of material system from
nonequilibrium state to locally-equilibrium state.

One can see  that the transition of material system into locally
equilibrium state is accompanied by origination of physical
structures. Massless particles, charges, structures made up by eikonal
and potential surfaces, wave fronts, and so on are
examples of physical structures.

(Here it should pay attention to the fact that to every physical field 
it is assigned its own material system. At present the question of what 
material system is assigned to given physical field remains unsolved. 
As it was mentioned before, the thermodynamical, gas dynamical, 
cosmological systems, the systems of charged particles and others are 
examples of material systems. Maybe, the physical vacuum is such 
material system for elementary particles.)

Identical relation (11) holds the duality. The left-hand side of this
relation includes the differential, which specifies material system
and whose availability points to locally-equilibrium state of
material system. And the right-hand side includes the closed inexact
form, which is a characteristics of physical structures and
from which the physical fields are formatted.

Such a duality of identical relation (11) emphasizes its unique
nature. This relation demonstrates the connection between material
systems and physical fields. And this connection is given up by
the evolutionary process, which is described by nonidentical
evolutionary relation obtained from the equations of balance
conservation laws for material systems.

The duality of identical relation also explains the duality of
nonidentical evolutionary relation. On the one hand, evolutionary
relation describes the evolutionary process in material systems,
and on the other describes the process of generating physical fields.

The origination of physical structures in evolutionary process is
connected with discrete changes of measurable (inherent)
quantities of material system and reveals in material system as an
emergence of certain observable formations, which develop
spontaneously. Such formations and their manifestations are
fluctuations, turbulent pulsations, waves, vortices, and others.
The spontaneous emergence of observable formations in material
system explains the mechanism of such processes like the
development of instability, the advent of vorticity [10], the
origination of turbulence.

The observed formation and the physical structure are not identical
objects. If the wave be such a formation, the element of wave front
made up the physical structure at its motion.
The structures of physical fields and the formations of material systems
observed are a manifestation of the same phenomena. The light is
an example of such a duality. The light manifests itself in the form
of massless particle (photon) and of a wave.

(This duality also explains a distinction in studying the same phenomena
in material systems and physical fields. In the physics of continuous
media (material systems) the interest is expressed
in generalized solutions to the equations of balance conservation laws.
These are solutions that describe the formations in material media
observed. The investigation of relevant physical structures is carried
out using the field theory equations.)

\bigskip
Thus, it has been shown that there exists a connection of physical 
fields with material systems. Material systems generate physical 
structures that made up physical fields. In this case the process of 
generation is controlled by conservation laws for material systems. 
Such controlling role of balance conservation laws in evolutionary 
processes is explained by noncommutativity of conservation laws.

[{\it The noncommutativity of balance conservation laws
and their governing role in evolutionary processes, that are
accompanied by emerging  physical structures, practically
have not been taken into account in the explicit form  anywhere. The
mathematical apparatus of evolutionary differential forms enables one
to take into account and describe these points.}]

Since physical fields are connected with material systems, it
seems natural to assume it has to be a connection between field
theories (those that describe physical fields) and the equations
for material systems (the equations of balance conservation laws
for energy, linear momentum, angular momentum, and mass as well as
the analog of such laws for time, which takes into account the
noncommutativity of time and energy of material systems). In
particular, it has to exist a connection of the field theory
equations with the equations for material systems. And this connection 
has to be described by closed exterior and evolutionary
skew-symmetric forms, which correspond to conservation laws for
physical fields and material systems.

\section{Peculiarities of field theories}

\subsection{Substantiation of basic principles of existing field 
theories}

In subsection 2.2 it has been shown that the properties of closed 
exterior forms, which correspond to conservation laws for physical 
fields, lie at the basis of existing field theories. The results 
obtained from the analysis of the equations of conservation laws
{\it for material systems} allows to substantiate these positions. 

It should be underlined the following results, which give the answer to
the questions posed in subsection 2.2.

a). Closed exterior forms on which properties the theories describing
physical fields are based, are obtained from
evolutionary forms entering into nonidentical relation derived from
the equations of balance conservation laws for material systems.

b). As it was shown above the degree of closed exterior forms, which 
plays a role of classification parameter for physical fields, is 
a degree of closed inexact exterior forms realized from evolutionary 
form in nonidentical relation following from the equations of 
noncommutative balance conservation laws for material systems.
From evolutionary forms, whose degree $p$ relates to the number of 
interacting balance conservation laws and can take the values 
$0, 1, 2, 3$, the closed (inexact) forms of degrees $k$ realize and  
can take the values $p, p-1, ..., 0$.

From this one can see that the degree of closed exterior forms, 
which plays a role of parameter for united  field theory. is 
connected with the number of the equations of interacting 
noncommutative balance conservation laws and can change from $k=0$ to 
$k=3$.

c). The external symmetries of the equations of field theory are
symmetries of closed dual forms. Such symmetries
are conditioned by degrees of freedom of material system
(translational, rotational, oscillatory and so on). Hence, the
external symmetries of the equations of field theory are also
conditioned by degrees of freedom of material system.

The gauge, i.e. internal, symmetries of field theory
(corresponding to gauge transformations) are those of closed
exterior forms. The symmetries of closed exterior forms are
obtained from the evolutionary form coefficients and
therefore is connected with the characteristics of material system.

As the result, the symmetries of closed exterior forms and,
consequently, the interior symmetries of field theory are defined
by the characteristics of material system.

d). The gauge transformations in field theories, which are
nondegenerate transformations of closed exterior forms, are
connected with degenerate transformations of evolutionary forms
obtained from the equations of conservation laws for material
systems. (Nondegenerate transformations of closed exterior forms
is a transition on integrable manifold from any closed inexact
form to another closed inexact form. However, every of these
closed inexact forms is obtained from evolutionary form defined on
nonintegrable manifold by using degenerate transformation.)

e). Since physical fields, as it has been shown, are formed up by
physical structures, this means that physical fields are discrete ones
rather then continuous. The discreteness of physical fields points to
the fact that field theories mast be quantum ones. This explains
the quantum character of field theories.

Characteristics of physical structures are connected with the 
characteristics of material systems. From this it follows that 
the constants of field theory must be connected with parameters 
of material systems.

\bigskip
The results that prove the basic principles of field theories were 
obtained while studying the equations of conservation laws for material 
systems. This points to the fact that the equations of field theory 
must be connected with the equations for material systems. 
Such a connection, which will be described in next subsection, clarify 
the specific features of the field theory equations.

\subsection{Connection between the equations of field theory and
the equations for material systems}

Before discussion of the equations of existing field theories it
should call attention to functional peculiarities of the equations
of mathematical physics and to nonidentical evolutionary relation.

\subsection*{Specific features of the equations of mathematical physics}

We will point out some peculiarities of the equations for material 
systems and those for field theories.

The equations for material systems are equations of balance
conservation laws of energy, linear momentum, angular momentum,
and mass. These are the equations of mechanics of continuous
media, as well as the equations of continuous medium physics,
which describe physical processes. In physics the interest is in
generalized solutions (which depend only on variables) that describe
observed quantities (rather then the process itself). To do this,
it is necessary to find the condition of integrability of the
system of differential equations. That is, it is necessary to
investigate the functional relation (nonidentical evolutionary
relation) obtained from differential equations.

The goal of continuous medium mechanics is to describe the process
of variation of continuous medium. In this case for solving
differential equations the numerical methods are commonly used
without analyzing the conditions of integrability of these
equations. This leads to that the solutions to equations depend
not only on variables but also on the path of integration.

As the result one has that in physics only generalized solutions
(which are functions only of variables) are considered and isn't
considered the solutions to original equations, which are not
generalized solutions, and in mechanics the generalized solutions
(for obtaining which one must investigate the integrability of 
equations) are not separated out. Such limited approach both in physics 
and in mechanics lead to nonclosure of relevant theories. Without 
finding generalized solutions in mechanics it is impossible to describe 
such processes like turbulence, emergence of vorticity and so on 
(i.e. the processes of emerging any formations that can be described by 
only generalized solutions). And without accounting for solutions that 
are not generalized ones in physics it is impossible to describe
the reason of obtaining generalize solutions.

Whereas the distinction between mechanics and physics of continuous 
medium consists in that both in mechanics and physics different types 
of solutions to the same system of equations are used, the distinction 
between physics of continuous medium and physics, which describes 
physical fields (i.e. field theory), is connected with use of 
distinct differential equations. 

The equations for material systems and the field theory equations 
are radically distinguished types of differential equations.

The fundamental difference between these two types is due to the
fact that the solutions to differential equations for material systems
have to describe physical {\it quantities}  (of material systems),
whereas the solutions to field theory equations have to describe 
physical {\it structures} (formatting physical fields). The basic 
element of the equations for material systems are {\it derivatives},
by integration of which one can obtain the desired functions
describing physical quantities. In contrast to that, the
properties of differential forms, that are, {\it differential
expressions and differentials } are used as the basis of field
theory equations. This is explained by the fact that the physical
structures, from which physical fields are made up, are described
by closed exterior forms, that is, by differentials.

Thus, it should to distinguish two types of equations of
mathematical physics: the differential equations for material
systems and the equations of field theory.
{\footnotesize (This explains a distinction in studying
the same phenomena in material systems and physical fields.
In the physics of continuous medium (material systems) the solutions to
equations (which are generalized solutions to equations of
balance conservation law) describe the observed formations in material 
media. The investigation of relevant physical structures is carried
out using the solutions to field theory equations.)}

It appears that these two types of equations are mutually connected.

It was shown that closed exterior forms, which describe
physical fields and, hence, must be the solutions to field theory
equations, follow from the nonidentical evolutionary relation, which is
derived from the equations for material systems. This means that there
exists a connection between field theory equations and the equations
for material systems. And this connection is realized with the help
of nonidentical evolutionary relation.

\subsection*{Functional peculiarities of nonidentical evolutionary
relation}

As it is shown (see formula (9)) the evolutionary
relation is derived from the equations of balance conservation
laws of energy, linear momentum, angular momentum, and mass and
in  general case (as it was shown) can be written as
$$
d\psi \,=\,\omega^p \eqno(9)
$$
where $\psi $ are {\it functionals} like wave-function, action
functional, entropy, the Pointing vector and others, $\omega^p $
is the evolutionary form of degree $p$, which depends on the
characteristics of material system and the external action to the
system, the form degree  $p$ takes the values $p\,=\,0,1,2,3$,
which are connected with the number of interacting noncommutative
balance conservation laws.

The nonidentical evolutionary relations are relations in differential
forms for {\it functionals}.

Such functionals like wave-function, action functional, entropy, and
others are used both in field theory and in the theories that describe
material systems (in mechanics and physics of continuous medium).
As it is seen from the analysis of nonidentical relation in section 4,
in the theories that describe material systems these functionals specify
the state of material system. And in field theory they describe
physical fields -- the field theory equations are those for such 
functionals. That is, such functionals possess a duality. This
duality of functionals and relevant nonidentical relations just allows
to disclose the connection between the field theory equations, which
describe physical fields, and the equations for material systems.

The equations for material systems (the equations of balance
conservation laws for energy, linear momentum, angular momentum,
and mass) are partial differential equations for desired functions
like the velocity of particles (elements), temperature, pressure
and density that correspond to physical quantities of material
systems (continuous media). The functionals like wave-function,
action functional, entropy and others and corresponding
nonidentical relations in continuous medium physics are used only
for analysis of integrability of differential equations, that is,
as a tool for obtaining the required solutions. {\footnotesize
[The analysis of integrability of differential equations is
necessary for to obtain the generalized solution that describes
measurable physical quantities of material system. The identity of
relation obtained from evolutionary equation means that original
equations for material system (the equations of conservation laws)
become consistent and integrable on pseudostructures. Identity of
the relation obtained from evolutionary relation means that the
original equations for material system become consistent and integrable 
on pseudostructures.
Pseudostructures made up integral surfaces (such as
characteristics, potential surfaces and others) on which the
desired quantities of material system (such as temperature,
pressure, density) become functions of only variables and don't
depend on the path of integration. These are generalized
solutions).]}

The field theory equations are those that describe physical
fields. Since physical fields are made up by physical structures,
which are described by closed exterior {\it inexact } forms, it is
obvious that solutions to the field theory equations must be closed
exterior forms, i.e. to be differentials. And such differentials, which
are closed exterior forms, can be obtained from nonidentical 
evolutionary relation for functionals. This means that in field theory 
the nonidentical evolutionary 
relations for functionals can play a role of field theory equation.
(As it will be shown later, in fact all equations of existing
field theories are the analog to such relation or its differential
or tensor representation.)

It is seen that the nonidentical evolutionary relation possesses 
the duality: this relation is used both in mechanics and physics of 
continuous medium and in field theory. The nonidentical relation unifies 
existing theories, namely, the field theory and the theory of
continuous medium, and make them closed ones.

\subsection*{Role of nonidentical evolutionary relation as the equation
of general field theory}

If the nonidentical evolutionary relation be regarded as the equation
for deriving identical relation with including closed forms (describing
physical structures desired), one can see that there is a correspondence
between such evolutionary relation and the equations of existing field
theories.

As it was already pointed out, at the basis of all existing field
theories there lie the properties of closed exterior forms that
correspond to conservation laws for physical fields. Since the
conservation laws are described by closed inexact exterior and dual
forms, it is evident that relevant closed exterior and dual forms
must be found from the field theory equations. One can ensure that
the identical relations, from which relevant closed exterior and dual
forms follow, are the solutions to equations of all existing field
theories.

The peculiarity of field theory equations consists in the fact that all
these equations have the form of relations. They can be relations in
differential forms or in the forms of their tensor or differential
(i.e. expressed in terms of derivatives) analogs.

The Einstein equation is a relation in differential forms. This equation
relates the differential of the first degree form (Einstein's tensor)
and the differential form of second degree, namely, the energy-momentum
tensor. (It should be noted that Einstein's equation is obtained from
differential form of third degree).

The Dirac equation relates  Dirac's {\it bra-} and {\it cket}- vectors,
which made up the differential forms of zero degree.

The Maxwell equations have the form of tensor relations.

Field and Schr\H{o}dinger's equations have the form of relations
expressed in terms of derivatives and their analogs.

All equations of existing field theories have the form of
nonidentical relations. Such peculiarity of the equations is
explained by the fact that only equations that have the form of
relations can have solutions being differentials rather then
functions. From such  equations it follows the identical
relations, from which the closed forms, which describe physical
structures, are found.

The identical relations, which include  closed exterior forms or their
tensor or differential analogs, are obtained from existing field theory
equations.

 By analyzing the field theory equations one can see that
from the field theory equations it follows such identical
relation like

1) the Dirac relations made up of Dirac's {\it bra-} and
{\it cket}- vectors, which are connected with closed exterior form of 
zero degree [11];

2) the Poincare invariant $ds\,=-\,H\,dt\,+\,p_j\,dq_j$, which is 
connected with closed exterior form of first degree;

3) the relations $d\theta^2=0$, $d^*\theta^2=0$ for closed exterior
forms of second degree obtained from Maxwell equations [4];

4) the Bianchi identities connected with skew-symmetric forms of third 
degree [12].

\bigskip
It turns out that all equations of existing field theories are in
essence relations that connect skew-symmetric forms or their
analogs. In this case one can see that the equations of field
theories have the form of relations for functionals such as wave
function (the relation corresponding to differential form of zero
degree), action functional (the relation corresponding to
differential form of first degree), the Pointing vector (the
relation corresponding to differential form of second degree).
The tensor functionals (such as the Riemann-Christoffel tensor,
the Ricci tensor), that correspond to Einstein's
equation, are obtained from the relation connecting the
differential forms of third degree.

The nonidentical evolutionary relation derived from the equations for
material system unites the relations for all these functionals. This is, 
all equations of field theories are an analog of nonidentical 
evolutionary relation. From this it follows that the nonidentical 
evolutionary relation can play a role of the equation of general field 
theory that discloses common properties and peculiarities of existing 
equations of field theory.

The correspondence between the equations of existing field theories
and the nonidentical evolutionary relation has the mathematical and
physical meaning. Firstly, this discloses the internal connection
between all physical theories. And secondly, this shows what has to lie
at the basis of general field theory, i.e. solves the problem of 
causality.

1. Encyclopedic dictionary of physical sciences. -Moscow, Sov.~Encyc.,
1984 (in Russian).

2. Cartan E., Les Systemes Differentials Exterieus ef Leurs Application
Geometriques. -Paris, Hermann, 1945.

3. Schutz B.~F., Geometrical Methods of Mathematical Physics. Cambrige
University Press, Cambrige, 1982.

4. Wheeler J.~A., Neutrino, Gravitation and Geometry. Bologna, 1960.

5. Petrova L.~I., Role of exterior and evolutionary skew-symmetric 
differential forms in mathematical physics. 
http://arxiv.org/abs/math-ph/0510077, 2005.

6. Tolman R.~C., Relativity, Thermodynamics, and Cosmology. Clarendon Press,
Oxford,  UK, 1969.

7. Clark J.~F., Machesney ~M., The Dynamics of Real Gases. Butterworths,
London, 1964.

8. Dafermos C.~M. In "Nonlinear waves". Cornell University Press,
Ithaca-London, 1974.

9. Bott R., Tu L.~W., Differential Forms in Algebraic Topology.
Springer, NY, 1982.

10. Petrova L.~I., The mechanism of generation of physical structures. 
//Nonlinear Acoustics - Fundamentals and Applications 
(18th International Symposium on Nonlinear Acoustics, Stockholm, Sweden, 
2008), New York, American Institute of Physics (AIP), 2008, pp.151-154.

11. Dirac P.~A.~M., The Principles of Quantum Mechanics. Clarendon Press,
Oxford, UK, 1958.

12. Tonnelat M.-A., Les principles de la theorie electromagnetique
et la relativite. Masson, Paris, 1959.

\end{document}